# EPR Paradox, Quantum Nonlocality and Physical Reality.

M Kupczynski

Département de l'Informatique, UQO, Case postale 1250, succursale Hull, Gatineau. QC, Canada J8X 3X 7

marian.kupczynski@uqo.ca

**Abstract**. Eighty years ago Einstein, Podolsky and Rosen demonstrated that the instantaneous reduction of wave function, believed to describe completely a pair of entangled physical systems, led to EPR paradox. The paradox disappears in statistical interpretation of quantum mechanics (QM) according to which a wave function describes only an ensemble of identically prepared physical systems. QM predicts strong correlations between the outcomes of measurements performed on different members of EPR pairs in far-away locations. Searching for an intuitive explanation of these correlations John Bell analysed so called local realistic hidden variable models and proved that these models always satisfy Bell inequalities which are violated by the predictions of QM and by experimental data. Several different local models were constructed and inequalities proven. Some eminent physicists concluded that Nature is definitely nonlocal and that it is acting according a new law of nonlocal randomness. According to this law perfectly random, but strongly correlated events, can be produced at the same time at far away locations and a spatio–temporal, local and causal explanation of their occurrence cannot be given. We strongly disagree with this conclusion and in this paper we criticise various finite sample proofs of Bell and CHSH inequalities and so called Quantum Randi Challenges. We also show how one can win so called Bell's game without violating locality of Nature. Nonlocal randomness is inconsistent with local quantum field theory, with standard model in elementary particle physics and with causal laws and adaptive dynamics prevailing in the surrounding us world. The experimental violation of Bell-type inequalities does not prove the nonlocality of Nature but it only confirms the contextual character of quantum observables and gives a strong argument against the point of view according to which the experimental outcomes are produced in irreducible random way. In spite of the fact that we have no doubt that Bell type inequalities are violated we show that because the homogeneity of samples was not tested carefully enough the significance of violation cannot be trusted.

**1. Introduction.**
Eighty years ago Einstein, Podolsky and Rosen (EPR) [1] demonstrated that an instantaneous reduction of a wave function describing a couple of entangled physical systems led to so called EPR paradox.

The paradox disappears in a statistical interpretation of QM according to which a wave function describes only an ensemble of identically prepared physical systems [2-5]. Measured values of physical observables corresponding quantum probabilities depend strongly on experimental context. Quantum probabilities are not degrees of belief of some intelligent agents but are objective properties of physical phenomena and experiments. Whether these probabilities can be deduced from some more detailed description of quantum phenomena is an open question.

In order to reproduce quantum statistics "*sub-quantal*" descriptions have to introduce supplementary parameters which are called hidden variables. Fifty years ago John Bell [6,7], trying to explain strong correlations between spin polarization projections of two physical systems prepared in spin singlet state, analyzed a large class of so called local realistic hidden variable models (LRHV) and found that correlations predicted by these models obeyed Bell inequalities (BI) which were violated by some correlations predicted by QM. Five years later John Clauser et al. [8] derived CHSH inequalities which are particularly suited for experimental testing. Using LRHV or stochastic hidden variables models (SHV) several other inequalities were proven and shown to be violated by QM and by the data of several ingenious experiments [9-15].

In this paper we concentrate on spin polarization correlation experiments (SPCE) with photons but some of our conclusions apply also to recent experiment with electrons by Hensen at al. [15].

In SPCE a source is sending two correlated signals to distant polarization beam splitters (PBS) and detectors. The outcomes on far away detectors, registered in some carefully chosen time windows, are 1=polarization up, -1= polarization down and 0=no count. Outcomes seem to appear randomly but distant time series are correlated more strongly than it is permitted by LRHV and SHV. These correlations are consistent with the predictions of QM. One may only conclude that a sub-quantal description of quantum phenomena cannot be based on LRHV or SHV.

However a large majority of quantum information community inspired by papers of John Bell, adopting a particular interpretation of a quantum state and it's reduction, concluded that a local and a causal explanation of correlations observed in SPCE is impossible and that *Nature is nonlocal*.

The violation of Bell type inequalities is believed to imply a violation of local realism (LR) as defined par example by Richard Gill [16]: " *local realism= locality + realism, is closely related to causality*"..." *events have causes (realism); cause and effect are constrained by time and space (locality)*".

This conclusion is incorrect because the assumption of *local realism* or *Bell locality* used in different proofs it is not LR defined above.

Nevertheless *quantum nonlocality* is considered to be a mysterious property of Nature. Apparently a mystery has a seductive power and not only excellent fiction writers but also distinguished scientists allow themselves to over-exaggerate quantum paradoxes. Entangled "photon pairs" are compared to pairs of fair dices which in each trial give perfectly matching outcomes. This is of course impossible. <u>Rolling of fair dices produces random outcomes which cannot be perfectly correlated</u>.

In his recent book Nicolas Gisin [17] explains in a pedagogical way *quantum nonlocality* and reviews results of many experimental and theoretical papers on the subject. He claims that quantum correlations observed in SPCE cannot be explained by causes belonging to the common past: *"Nature does not satisfy the continuity principle ...Nature is nonlocal"*. He advocates a new law of Nature: *"We must accept... nonlocal randomness, an irreducible randomness that manifests itself in several widely separated places without propagating from one point of space to next"*. At the same time randomly produced events are strongly correlated creating what means inexplicable mystery. In [17] one finds also statements such as: *"A particle passes through two neighboring slits at the same time. Therefore, an electron is indeed both here and a meter to the right of here"*. <u>This is not what QM says!</u>

When promoting his new law of *nonlocal randomness* Gisin is not impressed by the fact that quantum field theories (QFT) and a standard model in particle physics are local theories. He seems to forget also that all biological phenomena point towards adaptive dynamics and local causality.

For example in their fall migration tiny birds Bar-tailed Godwit (Limosa lapponica baueri) fly from Alaska to New Zealand 11 000 km in about eight days over the open Pacific Ocean, without stopping to rest or refuel. How could it be possible if *nonlocal randomness* was a law of Nature?

The *nonlocal randomness* is not needed in order to understand long range correlations in SPCE. LRHV do not use LR defined above. They use the assumption of *counterfactual definiteness* (CFD) according to which values of quantum observables, including incompatible ones, are predetermined before a measurement and are recorded passively by measuring instruments [18].

In SHV models do not use the predetermination but assume that results of measurements in distant laboratories are obtained in irreducibly random way what destroys all non-trivial correlations created by a source.

If these assumptions are not used various Bell type inequalities cannot be proven [19-50] and their violations give neither information about the locality of Nature nor about the completeness of QM.

Various probabilistic models used to prove inequalities are not consistent with experimental protocols used in SPCE [37, 44]. If contextual character of quantum observables is properly taken into account correlations may be explained in intuitive way [28-33, 40-47]. Moreover many experiments in quantum optics and in neutron interferometry can be simulated event by event in a local and causal way [51-54].

Unfortunately it seems that these results are not understood or ignored by part of physical community. This is why in this paper we give a detailed critical analysis of some influential finite sample proofs of Bell type inequalities and we hope it will finally close the issue.

This paper is organized as follows.

In section 2 we recall EPR paradox and Bohm's spin version of it (EPR-B) and the explanation given by a statistical and contextual interpretation of QM.

In section 3 we explain long range correlations in SPCE and their dependence on how pairing of distant outcomes is done.

In section 4 we analyze finite sample proofs of CHSH and Bell given by Richard Gill [16, 55] and Sacha Vongher [56] and we discuss impossible quantum Randi challenges proposed by them

In section 5 we show how one may win so called Bell's game discussed in [17] without invoking *nonlocal randomness*.

In section 6 we show how long range correlations in SPCE can be explained in a local and causal way if contextual character of quantum observables is correctly taken into account.

In section 7 we present few results from our paper written with Hans De Raedt [57] showing that sample *homogeneity loophole* was not closed in experiments testing various Bell type inequalities.

In section 8 we present our point of view on Physical Reality and it's abstract description provided by QT.

The last section contains conclusions.

## 2. EPR paradox and statistical interpretation.

Let us start with few axioms of QM which were believed to be true before the publication of the EPR paper:

- A1: Any pure state of a physical system is described by a specific <u>unique</u> wave function $\Psi$.
- A2: Wave function reduction: any measurement causes a physical system to jump into one of eigenstates of the dynamical variable that is being measured. This eigenstate becomes a new wave function describing the system after the measurement.
- A3: A wave function $\Psi$ provides a <u>complete</u> description of a pure state of an <u>individual</u> physical system.

EPR considered two particular individual systems I+II in a pure quantum state, which interacted in the past, separated and evolved freely afterwards [1]. Using A2 they concluded that:

- A single measurement performed on one of the systems, for example on the system I, gives instantaneous knowledge of the wave function of the second system moving freely far away.
- By choosing two different incompatible observables to be measured on the system I it is possible to assign two different wave functions to the same physical reality (the second system after the interaction with the first).

Since the measurement performed in a distant location on the system I does not disturb in any way the system II thus according to A1 and A3 it should be described by a unique wave function not by two different wave functions. Moreover these wave functions are eigenstates of two non- commuting operators representing incompatible physical observables what allows to deduce indirectly the values of these incompatible physical observables for the system II without disturbing it in any way what contradicts Heisenberg uncertainty relations.

Bohr [58] promptly reacted to EPR paper and pointed out that it was not possible to assign two different wave functions to the same reality (the second system after the interaction with the first) since the different wave functions could be assigned to the system II only in two different incompatible experiments in which both systems were exposed to different influences before the measurement on the system I was performed.

Bohr's arguments show that different eigenfunction expansions of the same wave function $\Psi$ provide probabilistic predictions for the behavior of identically prepared physical systems in different mutually excluding (complementary) experimental contexts. Therefore $\Psi$ is not an attribute of a single couple of these systems but only a mathematical tool used to deduct statistical regularities in various experimental data.

As early as in 1936 Einstein [2] noticed that EPR paradox disappeared if purely statistical interpretation of QM was used: "*$\Psi$ function does not, in any sense, describe the state of one single physical system. Reduced wave functions describe different sub-ensembles of the systems*" [2]. The statistical interpretation has been effectively promoted by Leslie Ballentine [3, 4] : "*the habit of considering an individual particle to have its own wave function is hard to break ...though it has been demonstrated strictly incorrect*" .

According to the statistical contextual interpretation of QM (SCI) [4, 5, 31, 40, 41]:

1. A state vector $\Psi$ is not an attribute of a single electron, photon, trapped ion, quantum dot etc. A state vector $\Psi$ or a density matrix $\rho$ describe only an ensemble of identical state preparations of some physical systems
2. A mysterious wave function reduction is neither instantaneous nor non-local. In EPR experiment a state vector describing the system II obtained by the reduction of the entangled state of two physical system I+II describes only the sub-ensemble of systems II being the partners of those systems I for which a measurement of some observable gave the same specific outcome. Different sub- ensembles are described by different reduced state vectors.
3. A value of a physical observable, such as a spin projection, associated with a pure quantum ensemble and in this way with an individual physical system being its member, is not an attribute of the system revealed by a measuring apparatus; but is a characteristic of this ensemble created by its interaction with the measuring device [36].

Let us explain below EPR-B paradox [4, 59] in which a source produces pairs of particles in a spin singlet state and the explanation given by SCI.

According to the orthodox interpretation of QM each pair of photons is described by a state vector.

$$\Psi = (|+\rangle_P |-\rangle_P - |-\rangle_P |+\rangle_P)/\sqrt{2}. \tag{1}$$

where $|+\rangle_P$ and $|-\rangle_P$ are state vectors corresponding to photon states in which their spin is "up" or "down" in the direction **P** respectively. If we measure a spin projection of a photon I on the direction **P** we have an equal probability to obtain a result 1 or –1. If we obtain 1 a reduced state vector of the photon II is $|-\rangle_P$, if we obtain -1 a reduced state vector of the photon II is $|+\rangle_P$. By choosing a direction **P**, for the measurement to be performed on the photon I, when "the photons are in flight and far apart" we can assign different incompatible reduced state vectors to the same photon II. In other words: we can predict with certainty, and without in any way disturbing the second photon, that the **P**-component of the spin of the photon II must have the opposite value to the value of the measured **P**-component of the spin of the photon I. Therefore for any direction **P** the **P**-component of the spin of the photon II has unknown but predetermined value what contradicts QM.

The solution of EPR-B paradox given by SCI is simple the wave function reduction is not instantaneous and a reduced one particle state $|+\rangle_P$ describes only an ensemble of partners of the particles I which were found to have "spin down" by a spin polarization analyzer pointing in the direction **P**. For various directions **P** it is a different sub ensemble of particles II. Strong correlations between distant outcomes in EPR experiments are due to various conservation laws. More detailed discussion of EPR paradox and SCI may be found for example in [40].

In the next section we argue that long distance imperfect correlations in SPCE are due to a common history of signals hitting detectors.

**3. Long distance correlations in SPCE and their interpretation.**
QM predicts strong correlations between spin projections, which we represent here as random variables A and B, made in different directions characterized by angles $\theta_A$ and $\theta_B$ :
$$E(AB|\psi) = -\cos(\theta_A - \theta_B) \qquad (2)$$
It seems that QM predicts strict anti-correlation of all **P**-components of the spin when the coincidence measurements are performed in the same direction **P** on both photons in far-away locations. At the same time according to quantum mechanics for the photons which are prepared in a state $|+\rangle_P$ their **P`**-component of the spin has no a definite value if **P`** ≠ **P**.

Since the choice of a direction can be made when "the photons are in flight" it seems impossible to keep strict anti-correlations of spin projections on all possible directions unless QM is incomplete and spin projections on all possible directions are predetermined by a source and registered passively by measuring instruments. In this case we have a mixed statistical ensemble of photon pairs characterized by correlated and predetermined spin projections on all possible directions which are recognized by polarization analysers and registered by the detectors. This is the assumption of CDF which is used in LRHV.

Correlations which one may obtain using LRHV or SHV always satisfy Bell type inequalities [6,7]. These inequalities are for some directions violated by quantum correlations [2] and by experimental data [9-15].

This is why some scientists started to consider supra-luminal influences between members of the photon pairs or between distant experimental settings and when these speculations failed *quantum nonlocality* became a mystery of Nature.

Let us show below that QM predicts only imperfect correlations in SPCE and that it is possible to give a rational explanation these correlations.
  1. In SPCE a pulse from a laser hitting the nonlinear crystal produces two correlated signals propagating in opposite directions. Clicks on distant detectors are correlated.

2. Bohr strongly insisted on wholeness of quantum phenomena. We do not see pairs of photons when they are created and when they travel across the experimental set-up. We only register clicks on the detectors.
3. Quantum field theory tells us only that a photon is a one photon state of quantized electromagnetic field and how it can be used in quantum calculations without giving us any intuitive picture of photons or virtual photons.
4. QM mechanics does not predict strict anti-correlations [36, 39-41] because directions of spin polarization analyzers are not sharp and can only be defined by some small intervals $I_A$ and $I_B$ containing angles close to $\theta_A$ and $\theta_B$ respectively. Therefore even if detection efficiencies were perfect and if we dealt with a perfect singlet spin state the prediction of QM for measured expectation values is :

$$E(AB|\psi) = -\iint_{I_A I_B} \cos(\theta_1 - \theta_2) d\rho_A(\theta_1) d\rho_B(\theta_2) \qquad (3)$$

5. In order to estimate correlations one has to define specific time windows in distant locations and define a pairing of clicks observed. There are many cases when a click is observed only in one location or no clicks at all.
6. To explain outcomes of SPCE, instead of a singlet state, more complicated mixed quantum states have to be used and strict anti-correlations are not predicted even for sharp directions. For more details see a paper by Köfler et al [60].
7. LRHV models fail because they neglect a contextual character of QT: "*The measuring instruments must always be included as part of the physical situation from which our experience is obtained*" [61].

In SPCE we have two far away laboratories performing experiments *x* and *y* on two physical signals $S_1$ and $S_2$ produced by some source S. Outcomes of experiments (*x*, *y*) in each particular time window are (*a*, *b*) where a=±1,0 and b=±1,0. The outcomes form two ordered samples of data: $S_A = \{a_1, a_2, \ldots a_n \ldots\}$ and $S_B = \{b_1, b_2, \ldots b_n \ldots\}$. These outcomes are observations of two time series of random variables $\{A_1, A_2 \ldots A_n \ldots\}$ and $\{B_1, B_2 \ldots B_n \ldots\}$ called sampling distributions.
If all $A_i$ are independent and identically distributed (i.i.d) as some random variable *A* and all $B_i$ are i.i.d as some random variable *B* then outcomes of experiments *x* and *y* can be completely described by conditional generalized joint probability distributions (GJPD): $P(A=a, B=b | x, y, S_1, S_2)$.

GJPD, describing outcomes of distant experiments, have different properties than standard joint probability distributions of a multivariate random variable and they strongly depend on how pairing of distant outcomes is made [44].

For example let us define two pairings:
- Systematic pairing :     $S_{AB}(1k) = \{(a_1, b_k), (a_2, b_{k+1}), (a_3, b_{k+2}) \ldots\}$
- Random pairing :     $S_{AB}(R) = \{(a_s, b_t) | s \leq t$ and s and t are chosen at random$\}$ .

If $S_A = \{-11-11-11-1..\}$ and $S_B = \{1-11-11-11..\}$ then for k- odd paring we have perfect anti-correlations, for k-even pairing we have perfect correlations and for random pairing there are no correlations (Cov(A,B)=0).

In SPCE we have two synchronized clocks in both laboratories, time windows are chosen in function of them and an appropriate systematic pairing is used. There are several difficulties and uncertainties related to this procedure, called the coincidence-time loophole , which have to be overcome [62].

It is important to underline that:

- The correlations do not prove any causal relation between *x* and *y*.
- No communication or direct influence between *x* and *y* is needed for their existence.

Now let us describe a plausible contextual model, consistent with local causality, able to describe imperfect long range correlations observed in SPCE.

1. Signals $S_1$ and $S_2$, correlated at the source, when arriving to respective experimental settings x and y are described by supplementary parameters $\lambda_1 \in \Lambda_1$, $\lambda_2 \in \Lambda_2$ and $P(\lambda_1, \lambda_2)$.
2. A choice of a setting x has no influence on the measurements performed using a setting y in a distant location (no signaling).
3. Experimental settings in each location can be chosen randomly or in a systematic way and observed correlations do not depend on how the choice is made.
4. Measuring devices as perceived by incoming signals are described by supplementary parameters $\lambda_x \in \Lambda_x$, $\lambda_y \in \Lambda_y$, $P_x(\lambda_x)$ and $P_y(\lambda_y)$.
5. To preserve a partial memory of the correlations created by a source outcomes *a* and *b* have to be produced in a local and deterministic way in function of local supplementary parameters $\lambda=(\lambda_1,\lambda_2,\lambda_x,\lambda_y)$ describing signals and measuring devices in successive time windows. Namely a= $A_x(\lambda_1,\lambda_x)$ and b= $B_y(\lambda_2,\lambda_y)$ where $A_x$ and $B_y$ are functions equal ±1 or 0.

6. The correlations predicted by QM for SPCE may reproduced by expectation values:

$$E(AB|x,y) = \sum_{\lambda \in \Lambda_{xy}} P(\lambda) A_x(\lambda_1,\lambda_x) B_y(\lambda_2,\lambda_y) \qquad (4)$$

where $P(\lambda) = P(\lambda_1,\lambda_2) P_x(\lambda_x) P_y(\lambda_y)$ and $\Lambda_{xy} = \Lambda_1 \times \Lambda_2 \times \Lambda_x \times \Lambda_y$ depend on (x, y).

Let us draw a causal graph representing how successive results (a, b) are produced:

$$x \rightarrow \Lambda_x \rightarrow \lambda_x \rightarrow a \leftarrow \lambda_1 \leftarrow \Lambda_1 \leftarrow S_1 \leftarrow S \rightarrow S_2 \rightarrow \Lambda_2 \rightarrow \lambda_2 \rightarrow b \leftarrow \lambda_y \leftarrow \Lambda_y \leftarrow y \ . \qquad (5)$$

It is important to underline that a choice of experimental settings (x, y) does not depend on supplementary parameters $\lambda=(\lambda_1,\lambda_2,\lambda_x,\lambda_y)$ but of course parameters ($\lambda_x, \lambda_y$) strongly depend on the choice of the settings. The reasoning based on the symmetry which tries to prove that the assumption of a free will implies that $\lambda$ do not depend on the choice of settings is simply incorrect.

Let us note that it is not needed to evoke *nonlocal randomness* and *quantum magic* to give an intuitive causal interpretation of Hansen et al. [15] experiment. We base our discussion below on a pedagogical description of this experiment given by Howard Wiseman [63]. Alice and Bob in successive time windows create entangled states of their electrons with photons. Photons are sent to Juanita's laboratory. Alice and Bob randomly choose setting for measurements of their respective electrons. They obtain their measurement outcomes and Juanita performs a joint measurement of the photons sent by Alice and Bob. If Juanita registers undistinguishable photons it means that the electrons in distant locations were prepared in particular physical states in which the outcomes of the measurement of their spin polarisation projections are correlated. In this experiment the cause of the correlations is not a partial memory of a common past or instantaneous communication from Juanita but similar physical conditions created in far-away locations when successful measurements were performed. The decisions which measurement outcomes are post selected to estimate correlations depend on rare positive Juanita's results.

A detailed discussion of the intimate relation of the experimental protocols and probabilistic models and why (4) cannot be derived by partial integration from some larger common probability space was given in [44]. Let us mention here an important paper by Andrei Khrennikov [33] who constructed a rigorous Kolmogorov model for SPCE in which CHSH could not be proven.

Arguments against Bell type inequalities based on probabilistic models and *contextuality* seem to be not well understood therefore in the next section we examine some influential finite sample proofs of Bell and CHSH inequalities. We show that they contain flaws and use assumptions not valid for SPCE.

**4. Finite sample proofs of Bell type inequalities and quantum Randi challenges.**

Finite samples due to statistical fluctuations may violate Bell and CHSH inequalities even if a corresponding probabilistic model never violates them. Assuming particular experimental protocols Richard Gill [16, 55] found probabilistic bounds on possible violations of CHSH in function of a sample size N.

In his first influential paper [16] he studies an experimental set-up of five computers: O, x, y, $R_A$ and $R_B$.

- The computer O, called a source, sends two correlated messages: strings of the length N containing approximately 50% of 0s and 1s.
- The computers x and y are " measurement stations" producing the outputs ±1 in function of the messages received. The produced strings contain approximately 50% of -1 and 1.
- The computers $R_A$ and $R_B$, called randomizers, send randomly (as they were results of independent fair coin tosses) setting labels "1" or "2" for each pair of outcomes produced by x and y. <u>This is the only source of randomness</u>.
- Regrouping the pairs of outputs corresponding to four possible choices of pairs of labels: (1,1), (1,2), (2,1) and (2,2) we obtain 4 samples of sizes ≈ N/4 formed as by a random pairing $S_{AB}(R)$ of the distant outcomes ±1 produced by x and y. As we saw such pairing destroys pre-existing correlations between messages as in SHV [44].

Using this idealized model of SPCE he proves that a probability, of finding the violation of CHSH, so large as predicted by QM and found in the experiment of Weihs et al. [11], is smaller than $10^{-32}$.

The protocol described above can be modified in a way that a sample from some joint probability distribution of the outcomes (±1, ±1; ±1, ±1) corresponding to 4 possible experimental settings is prepared. From this sample, using the labels sent by randomizers, marginal samples for particular settings are extracted.

This protocol is described more in detail in [55]. An idealized SPCE with N subsequent " *photon pairs*" is analyzed. There is no losses and 4 possible experimental settings are chosen in a random way.

For clarity of the argument we replace N by 4N. Possible outcomes ±1 are values of 4 random variables A, A', B, B'. Assuming that "*measurements which were not done also have outcomes; actual and potential measurement outcomes which are independent of the measurement settings actually used by all the parties*" Gill describes 4N subsequent pairs using 4N x 4 spreadsheet of numbers ±1. The rows are labelled by an index j = 1,2…,4N and columns by A, A', B, and B'.

This spreadsheet defines a random sample of size 4N drawn from some joint probability distribution of 4 random variables (A,A',B,B') and marginal expectation values E(AB) values can be estimated by $\langle AB \rangle = \frac{1}{4N}\sum_{j=1}^{4N} A_j B_j$. Similarly one finds estimates $\langle A'B \rangle$, $\langle AB' \rangle$ and $\langle A'B' \rangle$.

Since for any row we have $A_jB_j + A_jB'_j + A'_jB_j – A'_jB'_j = \pm 2$ thus for each value of N we obtain CHSH inequality:

$$\langle AB \rangle + \langle AB' \rangle + \langle A'B \rangle - \langle A'B' \rangle \leq 2 . \tag{6}$$

In SPCE settings are chosen at random thus Gill constructs finite samples of expected size N for each experimental setting in the following way: " *Suppose that for each row of the spreadsheet, two fair coins are tossed independently of one another, independently over all the rows. Suppose that depending on the outcomes of the two coins, we either get to see the value of A or A', and either the*

*value of B or B'. We can therefore determine the value of just one of the four products AB, AB', A'B, and A'B', each with equal probability 1/4 , for each row of the table. Denote by <AB>obs the average of the observed products of A and B ("undefined" if the sample size is zero). Define <AB'>obs ,<A'B>obs and <A'B'>obs similarly . When N is large one would expect <AB>obs to be close to <AB> and the same for the other three averages of observed products. <u>Hence the equation (6) should remain approximately true when we replace the averages of the four products over all 4N rows with the averages of the four products in each of four disjoint subsamples of expected size N each</u>*".

Following this construction Gill finds some probabilistic bound on the violation of (6) and makes a conjecture:

$$\Pr\left(\langle AB\rangle_{obs} + \langle AB'\rangle_{obs} + \langle A'B\rangle_{obs} - \langle A'B'\rangle_{obs} \geq 2\right) \leq \frac{1}{2} \qquad (7)$$

He proposes also a Quantum Randi Challenge (QRC): "*Construct 4Nx4 spreadsheet deduce from them 4 marginal samples of the expected size N, find the expectation values :* $\langle AB\rangle_{obs}$, $\langle A'B\rangle_{obs}$, $\langle AB'\rangle_{obs}$ *and* $\langle A'B'\rangle_{obs}$ *and check the CHSH inequality . If the program reproducibly, repeatedly (significantly more than half the time) violates CHSH, then the creator has created a classical physical system which systematically violates the CHSH inequalities, thereby disproving Bell's theorem.*

Of course QRC is impossible since finite samples of expected size N, extracted as above , from a counterfactual 4Nx4 spreadsheet may not, as it was proven above, violate CHSH significantly and repeatedly more than half the time. It does not mean that CHSH may not be consistently violated by finite experimental samples or samples generated using a specific local contextual model (4).

In SPCE 4Nx4 spreadsheets do not exist. The outcomes are not predetermined by a source and in some time-windows no counts or only single counts are detected. Thus random variables A, A', B and B' take 3 values ±1 and 0 and a post selection is needed in order to extract interesting finite samples containing the outcomes obtained when a particular experimental settings are used. The correlations found using different incompatible experimental settings do not need to satisfy Bell , CHSH or CH inequalities [20-50].

CFD is also used by Sacha Vongher [56]. He considers 800 tennis balls with instructions written on them. Each ball results in a measurement 0 or 1 according to the angle it encounters and instructions (HV) it carries. Angles at the measuring stations and HV may change randomly for each pair. Vongher uses angles α=aπ/8 (a=0 or 3) and β=bπ/8 (b=0 or 2). The instructions on each ball tell what to output for any choice of the angles it encounters: ($A_0$, $A_3$) and ($B_0$, $B_2$) for Alice's and Bob's balls respectively. For each pair of balls we can fill one row of the 800 x 4 counterfactual spreadsheet (discussed by Gill) but now instead of -1 we input 0. Vongher adds an additional constraint , strict anti-correlation for a=b=0, what reduces degrees of freedom to 3 and he chooses for the remaining independent variables : $A_3$, $B_0$ and $B_2$. The different settings are labelled by d=|b-a| and with $N_{total}$=800 , $N_d$ ≈200 for d=0,1,2,3. The outcomes are counted by 8 counters $N_d(X)$ where X=E( equal) or X=U (unequal). In his model $N_0(E)$=0 and $N_0(U)$ ≈200 because of imposed strict anti-correlations . Using this notation and his 800x4 spreadsheet he proves <u>nicely</u> a finite version of Bell inequality:

$$N_1(U) \leq N_2(E) + N_3(U) \qquad (8)$$

He simulates his experiment with tennis balls running a computer program 1000 times. The results reveal statistical fluctuations present in any finite samples and are very instructive.

- No violation of Bell and CHSH inequalities was observed for 800 pairs.
- For a second model , in which strict anti-correlations for d=0 were kept but some pairs were not prepared , Bell inequality (8) and CHSH were violated 50% of the time in 1000 runs.
- For a third model, in which anti-correlations occurred only 87% of time , Bell inequality was violated 87% of time and CHSH still only 50% of time.

- For the simulations based on the idealized quantum model (2) Bell inequality was violated 91% and CHSH was violated 99% of time in 1000 runs.

QRC proposed by Vongher is the following . Write a computer program preserving the idea of CDF and strict anti-correlations for d=0, simulate 1000 finite samples and show that Bell and CHSH inequalities are violated so consistently as by samples generated using quantum model. Since the finite samples constructed Gill and Vongher are similar to those drawn from the populations described by probabilistic models for which Bell , CHSH and Eberhard inequalities are satisfied their QRC are impossible.

Counterfactual spreadsheets of Gill and Vongher have nothing to do with spreadsheets containing outcomes of SPCE. As we explained in the discussion following the equation (3) strict anti-correlations are neither predicted by QM nor observed in SPCE. "Photons" are neither tennis balls with the instruction written on them (LRHV) nor perfect dices (SHV).

A strong argument against nonlocal randomness was given by Hans de Raedt and Kristel Michielsen et al. [51-54] who simulated event by event in a local and causal way several experiments in quantum optics and in neutron interferometry.

One could also try computer simulations based on a particular contextual model (4) but it was not done.

**5. How to win Bell's game without nonlocal randomness.**

In order to explain in a pedagogical way *quantum nonlocality* Nicolas Gisin discusses a particular Bell's game [17].

Alice and Bob have two identical boxes each equipped with a joystick and a screen. If the joystick is pushed to the left or to the right from a vertical (neutral position) a result appears on a screen. The results are binary: 0 or 1. Alice and Bob synchronize their watches and move some distance apart. Starting at 9am every minute they push joysticks on their boxes and record joystick positions and the results on the screen. Alice does not know the choice of the position of the joystick made by Bob and vice versa.

Let x=0 or 1 denote positions of the joystick on the Alice's box and a=0 or 1 displayed results. Let y=0 or 1 denote positions of the joystick on the Bob's box and b=0 or 1 displayed results.

The rules of Bell's game are the following:
- Settings (positions) *x* and *y* are chosen randomly.
- Outcomes *a* and *b* are determined locally in function of *x* and *y*.
- If $(x, y) = (1, 1)$ and $a \neq b$ → 1 point gained.
- If $(x, y) \neq (1, 1)$ and $a=b$ → 1 point gained.
- Otherwise no point is gained.
- The game is won if the average score is greater than 3.

These rules can be summarized by a simple equation: $[a +b]_2 = x\, y$ ( a sum modulo 2 of *a* and *b* is equal to a normal product of *x* and *y* ). A point is gained if the equation is obeyed. A claim is made that *S* is always smaller than 3 thus the winning of Bell's game is impossible.

Let us analyze a proof given in [17]. Since the results *a* and *b* are determined locally, in function of *x* and *y,* thus boxes at each trial use one of 4 possible local programs which are denoted *i* and *j*.
- i=1 : a=0 for all x    j=1: b=0 for all y
- i=2 : a=1 for all x    j=2 : b=1 for all y
- i=3 : a = x            j=3 : b= y
- i=4 : a = 1-x          j=4 : b= 1-y

We have 16 combinations of programs (*i*, *j*). Programs can change at each minute. Programs (*i*, *j*) determine outcomes (*a*, *b*) for settings (*x*, *y*). Gisin displays all possible outcomes for 16 possible pairs of programs (*i*, *j*) in a 16 row table. In Table 1 we reproduce the first row of Gisin's table using slightly different notation.

**Table 1.** Counterfactual calculation of scores of Bell's game in [17].

| (i, j) | (a,b) for (0,0) | (a,b) for (0,1) | (a,b) or (1,0) | (a, b) for (1,1) | S |
|---|---|---|---|---|---|
| (1,1) | (0, 0) | (0, 0) | (0, 0) | (0, 0) | 3 |

For each of 3 first settings $a=b$ thus 1 point is gained, no point is gained for the setting (1, 1) for a total score $S=3$. This is the maximal possible score possible. Another possible score is $S=1$, for example if $(i, j)=(1,2)$ and $a=0$ and $b=1$ for all possible settings. Thus the average $<S>$ of scores reported in any set of the lines is necessarily smaller than 3.

The reasoning given above is counterfactual, misleading and incorrect since programs chosen on each side may change every minute. In Table 1 it is assumed that the same couple (1, 1) of programs is used for all possible 4 settings chosen one after another. This is not what is necessarily happening, minute after minute, during Bell's game thus one cannot use the scores of each line in order to estimate $<S>$ for the game.

We display in Table 2 below possible scores in 4 consecutive minutes of Bell's game

**Table 2.** Calculation of scores in 4 consecutive minutes of Bell's game.

| Time | (i, j) | (x, y) | (a, b) | S |
|---|---|---|---|---|
| 1 | (1,1) | (0,0) | (0,0) | 1 |
| 2 | (2,2) | (0,1) | (1,1) | 1 |
| 3 | (4,3) | (1,1) | (0,1) | 1 |
| 4 | (3,4) | (1,0) | (1,1) | 1 |

Even if there are no common causes responsible for a particular choice of protocols one can, for a particular finite sample, obtain $<S> \approx 4$ consistent with no-signalling. If $(i, j)$ and $(x, y)$ were chosen at random we would expect to obtain $<S> \approx 2$ in a long run. Therefore it is clear that in order to win Bell's game consistently the outcomes are neither predetermined as in Table 1 nor produced randomly.

The experiment described above is called Bell's game because it resembles an idealised SPCE for which QM, using (2), predicts the average score $<S> \approx 3.41$. Let us notice that if the outcomes for each setting are predetermined the experimental protocol of Bell's game is similar to the protocol used by Vongher [56] in which $(i, j)$ are instructions written on pairs of tennis balls.

If we reject *nonlocal randomness* we have to use a principle of a common cause, local causality and contextuality in order to explain how a choice of particular protocols $(i, j)$ may depend on settings chosen randomly in distant locations.

We follow the same logic which led us to equations (4) and (5). Every minute settings are chosen randomly and their microstates are described by two local parameters ($\lambda_x, \lambda_y$) drawn from $\Lambda_x \times \Lambda_y$. Two correlated signals, described by parameters ($\lambda_1, \lambda_2$) drawn from $\Lambda_1 \times \Lambda_2$, arrive to the Alice's and Bob's boxes. The outcomes $(a, b)=(i(x), j(y))$ where $i= f(\lambda_1, \lambda_x)$ and $j=g(\lambda_2, \lambda_y)$. Thus protocols on the distant machines are correlated and depend on microstates of the settings.

Calculations in the equation $[a +b]_2=x\ y$ cannot be done locally what means that neither Alice nor Bob can find both values $a$ and $b$ having only their local information. It does not mean that Nature does not satisfy the continuity principle and is nonlocal. The outcomes $(a, b)$ are not results of magical nonlocal calculations but they are produced in locally causal interactions of correlated signals with experimental devices in different experimental contexts.

## 6. Sample homogeneity loophole and violation of Bell type inequalities.

Bell -CHCH-CH inequalities cannot be proven if a contextual character of physical observables is correctly taken into account [22-47]. Therefore it is not strange for us that they are violated in SPCE and also in experiments in different domains of science [31, 32] and even in classical mechanics [42].

After several ingenious experiments and after closing several experimental loopholes [9-15] there seems to be strong experimental evidence that Bell type inequalities are violated in Nature.

However there are two problems troubling us. Both of them are related to finite statistics.

1. As Vongher demonstrated when <u>a particular local realistic model</u> was used to generate 1000 samples of size 800 it turned out that Bell inequality was violated 87% of time and CHSH 50% of time. To compare in Weihs et al. [11] two long runs were analysed and in only one a significant violation of CHSH was observed. In Giustina et al. [13] and in Christenson et al. [14] only one long run was analysed and the violation found. In Hensen at al. [15] we have only <u>one sample containing 245 data items</u>.
2. Statistical analysis made in these experiments assumed that studied data are simple random samples. Recently we reported with Hans de Raedt [57] a dramatic breakdown of statistical inference due to sample inhomogeneity. Therefore if sample homogeneity is not checked results of significance tests may not be trusted and we say that sample *homogeneity loophole* (SHL) was not closed [64].

To show how detrimental SHL can be we quote one example from our paper [57]. We simulated a random experiment in which a measuring device, operating according to some specific internal protocol, was outputting one of 6 possible discrete values. We generated 100 runs (each run containing $10^5$ data items). Using these large samples we made a significance test of a null hypothesis $H_0$: 1-B ≥ 0. When three runs 25, 50 and 75 were used the inequality was violated for each run by more than 2000 SEM (standard error of the mean) and one could with great confidence reject the null hypothesis. When we performed the average over 100 runs ($10^7$ data items) 1-B =+0.95 SEM and of course the null hypothesis could not be rejected. The reason was that samples produced by our device were not homogeneous.

Let us now comment on the experiment of Giustina et al [13]. In this experiment a source is sending polarization-entangled photons which after passing by one of four possible polarization measuring settings ($\alpha_i$, $\beta_j$) are sent to two detectors one operated by Alice and another by Bob. The clicks on the detectors are registered and the coincidence counts determined.

Each setting defines a different random experiment and in order to check Eberhard's inequality one estimates a value of a random variable $J$. This variable is a particular combination of values of 6 random variables (4 coincidence counts and 2 single counts) deduced from the data gathered in all four experimental settings. If CFD was true $J$ should be always positive thus the null hypothesis tested is $H_0$: $J \geq 0$.

The data gathered during 300 seconds of recording per setting were divided into 30 bins and a sample S of 30 different $J$-values was obtained. From this sample the value of the mean $<J>$ together with its standard mean error SEM= $s/n^{0.5}$ (n=30, s=sample standard deviation) were estimated and 67σ (67 SEM) violation of Eberhard's inequality was reported.

Usually we use this terminology if we are convinced that the central limit theorem can be used for a finite sample and $<J>$ is normally distributed. Since the normality of this distribution cannot be proven Khrennikov et al. [65] used Chebyshev inequality and concluded that the null hypothesis can be still rejected at the confidence level of 99.95%.

However this null hypothesis test is based on only one (1) sample containing 30 data items and the conditions of a simple random sample, were not tested. Since the whole set of data contains the outcomes from 4 different random experiments of course it would be surprising if it was homogeneous. A sample of observed values of J could only be homogeneous if the data sets obtained for each fixed setting were homogeneous and it was not tested carefully enough.

Giustina et al. [13] closed the " fair-sampling loophole". Köfler et al. [61] showed that the experiment was immune to the "production-rate loophole" and that the results were consistent with quantum theory. Larsson et al. [62] proved that the experiment was not vulnerable to the "coincidence-time loophole". It is clear that it is an excellent experiment performed with a great scrutiny. Nevertheless unless the additional sample homogeneity tests are performed, SHL is not closed and its impact on conclusions of the significance tests is unknown.

In a recent paper Andrei Khrennikov [66] also pointed out that the statistical analysis of data, which was reported as violating Bell's inequality, suffered of a number of problems.

One could wrongly understand that we do not believe in the experimental evidence of the violation of Bell type inequalities. As we already told above it is just the opposite.

**7. Physical Reality.**

All our science is built on the assumption that there exists an objective external world governed by some laws of Nature which we want to discover and to harness.

In Physics we construct idealized mathematical models in order to explain in qualitative and quantitative way various phenomena which we observe or we create in our laboratories [67-70].

Our perceptions are biased by our senses and by our brain and depend on the location from where they are made. For example motions of the planets are complicated when observed from the Earth and it took many centuries to explain them in a simple way using the heliocentric system and Newton's equations of motion. Stars and planets are perceived by a naked eye as clinking points on a sky. These images are created by our brain when the light hits the retina of our eyes but of course there is something real behind the scenes causing these perceptions, The stars and planets existed before the life existed on the Earth thus statements such as: A Moon does not exist if we don't look at it, made in order to impress a general public, are incorrect and misleading.

Similarly in quantum phenomena which we observe and create there should be something behind the scenes which is causally responsible for outcomes we register. In our opinion quantum probabilities neither correspond to irreducible propensities of individual physical systems nor to beliefs of some human agents but they are objective properties of quantum phenomena as a whole [4, 5, 31, 37, 67, 68, 70]. Necessity of probabilistic description is due to the lack of control on what is going behind the scenes.

In contrast to classical physics measuring instruments do not register, in general, pre-existing values of physical observables characterizing physical systems. The values of physical observables are obtained as a result of an incontrollable interaction of a physical system with a measuring device and any attempt to provide more details about what is going on behind the scenes has to include a description of the microscopic state of the device in the moment of the measurement .

Taking into account enormous successes of QM and QFT a more detailed unambiguous description of quantum phenomena does not seem to be needed or possible but telling that there is nothing behind the scenes is in our opinion naïve, unproven and unproductive. Even in macroscopic physics we are not able to describe in detail all observed phenomena. For example we neither can explain all ripples on a surface of a lake after passing of a boat nor motions of successive water droplets in a waterfall etc.

In quantum phenomenon we do not see "a lake and a boat" but it does not mean that they do not exist. Therefore the efforts to prove that QM is an emergent theory and to better understand its foundations may help to dismiss the *quantum magic* and the speculations that the Nature is non local.

If locality of interactions and causality were violated in the micro-world how could we have locally causal macro-world and how could we reconcile QM and QFT with General Relativity?

Let us finish this section with two remarks.
- QM and QFT describe statistical properties of quantum phenomena in a way consistent with Einsteinian locality.
- According to fathers of QM a question by which slit electron passes in two slit interference experiment is meaningless therefore a statement that an electron is at the same time here and a meter from here is nonsense. Similarly *quantum teleportation* is implementable experimental protocol but statements that Alice has two entangled photons, sends one of them to Bob and after makes a joint measurement of a remaining photon with a photon to be teleported are misleading. It simply cannot be done for three photons in question since QM and QFT tell nothing how to manipulate single photons in this fashion.

The quantum magic is born if exotic interpretations of QM are adopted or if naïve models of sub-quantum phenomena are constructed and examined.

At the same moment when we start evoke magic it is the end of physical explanation and we accept that we will never be able to understand. Fortunately some papers presented at this conference and at previous Emergent Quantum Mechanics conferences show that it is not necessarily true and that we can get some intuitive insight on what might be going under the scenes. Let us cite here as an example a paper by Grössing et al. [71].

**8. Conclusions**

Paradoxes are created if incorrect interpretation of QT is adopted or if inappropriate sub-quantal description of quantum phenomena is used. A statistical contextual interpretation of QM reconciles the ideas of Bohr and Einstein and is free of paradoxes.

QM, QFT and a standard model in elementary particle physics are consistent with Einsteinian locality. The proofs of various Bell type inequalities are based on CFD or on the assumption that the experimental outcomes are produced in irreducibly random way. If these assumptions are not valid Bell-type inequalities cannot be simply proven.

Therefore paraphrasing Howard Wiseman [63] the violation of these inequalities observed in several experiments hammers, in our opinion, the final nail in the coffin of *counterfactual definiteness* and *irreducible randomness*.

In order to explain strong long range correlations observed in SPCE we do not need to postulate a new law of Nature called *nonlocal randomness*. An intuitive, contextual and causal explanation may be given: signals are correlated by a source, keep partial memory of it when flying to distant laboratories and outcomes are produced in locally deterministic way in function of micro-states of the signals and of measuring devices at the moment of the measurement.

Therefore the violation of Bell type inequalities gives an additional argument in favor of a point of view that quantum probabilities might emerge from some underlying more detailed and causal description of quantum phenomena to be discovered

We have recently demonstrated with Hans de Raedt [57] that significance tests may dramatically break down if studied samples are not homogeneous. We do not doubt that various Bell type inequalities are violated in SPCE however the results of various significance tests may not be trusted since the sample homogeneity was not or could not be tested carefully enough.

Concerning the Physical Reality we strongly believe that there exists an external world whose existence does not depend whether it is observed or not. This external world is governed by laws of Nature which we try to discover. The quantum phenomena which we create depend on the devices used to probe this external world and on detailed contexts of our experiments. The information we get is contextual and complementary but quantum probabilities are the objective properties of quantum phenomena and not the beliefs of some human agents.

An additional argument in favor of the idea that QM might be an emergent theory would be a discovery that experimental time series of the data present some fine structures not predicted by the theory. It would not only prove that QM may not provide the most complete description of the individual physical systems but it would also prove that QM is not predictably complete [72, 73].

Let us finish this article with words of Einstein [2]:"*Is there really any physicist who believes that we shall never get any insight into these important changes in the single systems, in their structure and their causal connections…To believe this is logically possible without contradiction; but, it is so very contrary to my scientific instinct that I cannot forego the search for a more complete description*".

**Acknowledgments**
I would like to thank Gerhard Grössing and Fetzer Foundation for the invitation and their hospitality extended to me during this interesting conference.